\documentstyle[aps,preprint]{revtex}

\begin{document}
\draft
\preprint{US-FT/2-95}

\title{New no-scalar-hair theorem for black-holes}

\author{Alberto Saa}

\address{Departamento de F\'\i sica de Part\'\i culas\\
       Universidade de Santiago de Compostela\\
       15706 Santiago de Compostela, Spain}
\date{\today}
\maketitle

\begin{abstract}
A new no-hair theorem is formulated which rules out a very large
class of non-minimally coupled finite scalar dressing of an asymptotically
flat, static, and spherically symmetric black-hole. The proof is 
very simple and based in a covariant method for generating solutions for 
non-minimally coupled scalar fields  starting from the minimally 
coupled case. Such
method generalizes the Bekenstein method for conformal coupling 
and other recent ones. We also discuss the role of the finiteness
assumption for the scalar field.
\end{abstract}

\vspace{1cm}
\begin{flushleft}
PACS: 0420, 0450
\end{flushleft}

\newpage

\section{Introduction}

Black-hole solutions are very rigid in gravitational physics. We
know that the Schwarzschild solution is the only asymptotically
flat and spherically symmetric solution of the vacuum Einstein equations.
The no-hair conjecture\cite{RW}
states that the exterior region of a black-hole
admits only fields for which there is a geometrical Gauss-like law, as
electromagnetic fields for example. 
Early no-hair theorems excluding for the
exterior region of a black hole minimally coupled 
Klein-Gordon \cite{S}, massive vectors\cite{V}, 
and spinor\cite{Sp} fields have stressed the conjecture.

The problem of the existence of scalar hairs for black-holes
has received some attention
recently. Although we know that scalar fields are not elementary fields
in nature, they commonly arise in effective actions. 
In fact, some scalar actions have been considered recently in 
astrophysical contexts, see for instance \cite{B3}.
However,
with the conformally coupled case as the only exception\cite{xd,CK,Z}, 
only minimally
coupled scalar fields have been examined. 
In \cite{B1} it is presented a new theorem which rules out a multicomponent
scalar hair with non-quadratic Lagrangian, but with minimal coupling to 
gravity.
As it is stressed in 
 \cite{B1}, scalar fields effective actions are obtained by integrating
the functional integral of the elementary fields in nature over some of
the fields, and more complicated actions involving non-minimally coupling
should arise.

The purpose of the present work is to point toward the filling of this
gap by presenting a theorem that excludes finite scalar hairs of any
asymptotically flat, static, and spherically symmetric
black-hole solution of the system
described by the action
\begin{equation}
S[g,\phi] = \int d^4x\sqrt{-g} \left\{f(\phi) R - h(\phi) g^{\mu\nu}
\partial_\mu\phi\partial_\nu\phi\right\},
\label{s}
\end{equation}
with $f(\phi)$ and  $h(\phi)>0$. 
We adopt all the conventions of \cite{wi}. Many physically relevant theories
belong to the class described by (\ref{s}). Maybe the most popular non-minimal
coupling for the scalars fields corresponds to the choice 
$f(\phi) = 1 - \xi\phi^2$ and $h(\phi) =1$. 
The case $\xi=\frac{1}{6}$ corresponds to the conformal coupling case,
and the Bekenstein method\cite{b} allows 
us to construct its exact solutions from the solutions of the
minimally coupled case ($\xi=0$). A 
method for generating solutions for arbitrary
$\xi$ is presented in \cite{AWKP}.
The extension of Bekenstein method for $n$-dimensions ($n>3$) was 
obtained recently in
\cite{xd}, and used to study conformal scalar hairs\cite{xd,CK} 
and gravitational waves\cite{kk}.
Dilaton-like gravity is given by
$f(\phi) = \frac{1}{4}h(\phi) = e^{-2\phi}$. The general 
model of Bergman, Wagoner and Nordtved discussed in 
\cite{wi}  corresponds to the choice
$f(\phi) = \phi$ and $h(\phi) = \frac{\omega(\phi)}{\phi}$, from which 
Brans-Dicke theory is obtained from the limit $\omega$ constant.

The paper is organized as follows. In the section \ref{tt} we present
a covariant method for generating solutions for the system described 
by (\ref{s}). This method will be the central point for the formulation
of the theorem, which is presented in the same section. In the section
\ref{ttt}, we analyse as particular cases the Brans-Dicke theory
and one of its generalizations in order
to shed light in the role of the finiteness assumption for the scalar
field and its relation to naked singularities. The last section is devoted
to some concluding remarks, in particular a comparison between our results
and recent ones.

\section{The theorem}
\label{tt}
The proof of our theorem centers in a covariant method for generating
solutions for the Euler-Lagrange equations of (\ref{s}) starting from
the well known solutions for the minimally coupled case,
\begin{equation}
\bar{S}[\bar{g},\bar{\phi}] = \int d^4x\sqrt{-\bar{g}} 
\left\{\bar{R} - \bar{g}^{\mu\nu}
\partial_\mu\bar{\phi}\partial_\nu\bar{\phi}\right\}.
\label{sb}
\end{equation}
The method uses a conformal transformation, and generalizes the
Bekenstein one\cite{b} and the proposed in the Ref. \cite{AWKP}.
Such a kind of method has a long history, and the Ref. \cite{MS},
for instance, presents a good set of references on the subject.
A method of this type was also used in \cite{m} to show that the
action given by $\int d^4x\sqrt{-g}\left\{ 
F(\phi,R) - g^{\mu\nu}\partial_\mu\phi\partial_\nu\phi\right\}$ is
equivalent to an Einstein-Hilbert action plus minimally coupled
self-interacting scalar
fields, equivalent in the sense that there is a conformal transformation
and $\phi$-redefinition connecting them.

The Euler-Lagrange equations obtained from (\ref{s}) are
\begin{eqnarray}
\label{el1}
f(\phi)R_{\mu\nu} - h(\phi)\partial_\mu\phi\partial_\nu\phi -
D_\mu D_\nu f(\phi) 
- \frac{1}{2}g_{\mu\nu}\Box f(\phi) &=& 0, \\
2h(\phi)\Box\phi + 
h'(\phi)g^{\alpha\beta}\partial_\alpha\phi\partial_\beta\phi 
+ f'(\phi)R &=& 0,\nonumber
\end{eqnarray}
where the prime denotes derivation with respect to $\phi$. 
Equations (\ref{el1})
are clearly much more complicated than the Euler-Lagrange
equations derived from (\ref{sb}), namely
\begin{eqnarray}
\label{el2}
\bar{R}_{\mu\nu}  - 
\partial_\mu\bar{\phi}\partial_\nu\bar{\phi} &=& 0, \nonumber \\
\stackrel{-}{\Box}\bar{\phi} &=& 0. 
\end{eqnarray}
In order to realize how the solutions of (\ref{el1}) and (\ref{el2}) are
related, consider the conformal transformation
$g_{\mu\nu} = \Omega^2\bar{g}_{\mu\nu}.$
Under a conformal transformation, the
scalar of curvature transforms as
$R(\Omega^2\bar{g}_{\mu\nu}) = \Omega^{-2}\bar{R} -
6\Omega^{-3} \stackrel{-}{\Box}\Omega$, and with the choice
\begin{equation}
\label{ct}
f(\phi) = \Omega^{-2},
\end{equation}
one gets from (\ref{s})
\begin{equation}
S[\Omega^2\bar{g},\phi] = \int d^4x\sqrt{-\bar{g}} 
\left\{\bar{R} - \left(
\frac{3}{2}\left(\frac{d}{d\phi}\ln f(\phi)\right)^2 + 
\frac{h(\phi)}{f(\phi)}
\right)\bar{g}^{\mu\nu}\partial_\mu\phi\partial_\nu\phi\right\}.
\end{equation}
Now, defining $\bar{\phi}(\phi)$ as
\begin{equation}
\label{int}
\bar{\phi}(\phi) = \int_a^\phi d\xi \sqrt{
\frac{3}{2}\left(\frac{d}{d\xi}\ln f(\xi)\right)^2 + 
\frac{h(\xi)}{f(\xi)}},
\end{equation}
with arbitrary $a$, we get the desired result,
$S[\Omega^2\bar{g},\phi(\bar{\phi})] = 
\bar{S}[\bar{g},\bar{\phi}].$ Due to the assumption of $f$ and $h$ positive,
the right-handed side of (\ref{int}) is a monotonically increasing function
of $\phi$, what guarantees the existence of the inverse $\phi(\bar{\phi})$.
The constant $a$ is determined by the boundary conditions of
$\phi$ and $\bar{\phi}$.
Also, we have  that 
$\lim_{\bar{\phi}\rightarrow\infty} \phi(\bar{\phi}) = \infty$.

The transformation  given by eq. (\ref{ct}) and (\ref{int}), 
therefore, maps a solution $(g_{\mu\nu},\phi)$ of (\ref{el1}) 
to a solution $(\bar{g}_{\mu\nu},\bar{\phi})$ of (\ref{el2}). The
transformation is independent of any assumption of symmetries, and
in this sense is covariant. We can easily infer that the transformation
is one-to-one in general,
 in the sense that any solution of (\ref{el1}) is mapped
in an unique solution of (\ref{el2}). Also, the transformation
preserves symmetries, what means that if
$\bar{g}_{\mu\nu}$ admits a Killing vector $\xi$ such that
${\pounds}_{\xi}\bar{\phi} = 0$, then $\xi$ is also a Killing vector
of ${g}_{\mu\nu}$ and ${\pounds}_{\xi}{\phi} = 0$. From this, one concludes
if we know all solutions $(\bar{g}_{\mu\nu},\bar{\phi})$ with a given
symmetry we automatically know all $({g}_{\mu\nu},{\phi})$ with the
same symmetry. This is the base of the proof.

The general asymptotically flat, static, and spherically symmetric
 solution $(\bar{g}_{\mu\nu},\bar{\phi})$ 
of (\ref{el2}) is known (See \cite{X1} for some properties
of the solution and references). 
It is given by two-parameter $(\lambda,r_0)$ family of solutions
\begin{eqnarray}
\label{solu}
\bar{\phi} &=& \sqrt{2(1-\lambda^2)}\ln { \cal R}, \nonumber \\
ds^2 &=& \bar{g}_{\mu\nu} dx^\mu dx^\nu = 
-{\cal R}^{2\lambda}dt^2 + 
\left( 1-\frac{r_0^2}{r^2}\right)^2 {\cal R}^{-2\lambda}
\left(dr^2 + r^2 d\Omega^2 \right), 
\end{eqnarray}
where ${\cal R}=\frac{r-r_0}{r+r_0}$.  The parameter
$\lambda$ can take values in $[-1,1]$ in principle, 
but we neglect the negative range
because the solution will have a negative ADM mass\cite{X1}.
For $\lambda = 1$, the solution is the usual exterior vacuum Schwarzschild 
solution with the horizon at $r'_0=4r_0$, as one can check by using the 
coordinate transformation $r' = r\left(1+\frac{r_0}{r}\right)^2$.
For $0 \le \lambda < 1$, (\ref{solu}) does not represent
a black-hole due to that 
the surface $r=r_0$ is not a horizon, {\em i.e.} a regular null surface,
but it is instead
a naked singularity, as we can check, for instance, 
by calculating the scalar of
curvature 
\begin{equation}
\label{R}
\bar{R}= \frac{8r_0^2r^4}{(r+r_0)^{2(2+\lambda)}}
\times \frac{1-\lambda^2} {(r-r_0)^{2(2-\lambda)}}.
\end{equation}
In total accordance with the original scalar no-hair 
theorem\cite{S}, we see the only 
black-hole solution of (\ref{solu}) is that one
for which $\lambda=1$ and consequently $\phi=0$, {\em i.e.} the usual 
Schwarzschild solution.

Any asymptotically flat, static, and spherically symmetric
solution of (\ref{el1}) can be obtained from (\ref{solu}) by means of
the transformations (\ref{ct}) and  (\ref{int}). This provides us
with a two-parameters  family of $(g_{\mu\nu},\phi)$
solutions. The discussed properties of the transformation (\ref{int}) and the
expression for $\bar{\phi}(r)$ in (\ref{solu}) 
lead to the conclusion that the
only solution with $\phi$ finite in the surface $r=r_0$ is that one
for which $\phi$ is constant for $r>r_0$. In this case, (\ref{ct}) is only
a rigid scale transformation, and the solution $(g_{\mu\nu},\phi=a)$ is
the usual Schwarzschild solution. This is the desired result, which
we formulate for clearness as follows.

\noindent{\em Theorem.}
The only asymptotically flat, static, and spherically symmetric
exterior solution of the system governed by the action
$$
S = \int d^4x\sqrt{-g} \left\{f(\phi) R - h(\phi) g^{\mu\nu}
\partial_\mu\phi\partial_\nu\phi\right\},\ \ \ \ \ \ \ 
f(\phi), h(\phi)>0.  
$$
with $\phi$ everywhere finite is the Schwarzschild solution.

It is important to note that the used conformal transformation forbids 
that $f(\phi) \rightarrow\infty$ for any $r\ne r_0$.

Our approach can be extended in a straightforward way to other dimensions.
The transformations (\ref{ct}) and (\ref{int}) can be defined for 
any dimension $n>2$. They shall be replaced by 
\begin{eqnarray}
\label{intn}
f &=& \Omega^{2-n} \nonumber , \\
\bar{\phi}(\phi) &=& \int_a^\phi d\xi \sqrt{
\frac{n-1}{n-2}\left(\frac{d}{d\xi}\ln f(\xi)\right)^2 +
\frac{h(\xi)}{f(\xi)}}.
\end{eqnarray}
The general asymptotically flat, static, and spherically symmetric
solution for any space-time dimension
$n>3$ of (\ref{el2}) is known\cite{X1}. Its expression 
in isotropic coordinates is given by
\begin{eqnarray}
\label{solun}
\bar{\phi} &=& \sqrt{\frac{n-2}{n-3}(1-\lambda^2)}\ln {\cal R}_n, \nonumber \\
ds^2 &=& \bar{g}_{\mu\nu} dx^\mu dx^\nu =
-{\cal R}^{2\lambda}_ndt^2 + \left( 
1-\frac{r_0^{2n-6}}{r^{2n-6}}
\right)^\frac{2}{n-3} 
{\cal R}^{-\frac{2\lambda}{n-3}}_n 
\left(dr^2 + r^2 d\Omega^2 \right),
\end{eqnarray}
where ${\cal R}_n=\frac{r^{n-3}-r^{n-3}_0}{r^{n-3}+r^{n-3}_0}$ and $d\Omega$ 
denotes the metric of the unitary $(n-2)$ sphere. The behavior of
the solution (\ref{solun}) is similar to the four dimensional case.
The only true black-hole solution is the usual one $(\lambda = 1)$, due to 
the fact that the hyper-surface $r=r_0$ is not a regular one if
$\lambda \ne 1$, as one can see from the expression for the
scalar of curvature 
\begin{equation}
\bar{R} =  
\frac{4(n-2)(n-3)r_0^{2(n-3)}r^{2(n-4)}}
     {(r^{n-3}+r_0^{n-3})^{\frac{2(n-2+\lambda)}{n-3}}}
\times \frac{1-\lambda^2}{(r^{n-3}-r_0^{n-3})^{\frac{2(n-2-\lambda)}{n-3}}}. 
\end{equation}
By applying 
(\ref{intn}) and the same arguments used to the four dimensional case
we can extend our theorem for any space-time dimension $n>3$.

\section{An explicit example}
\label{ttt}
A closer look in an explicit example will help us to understand the role
of the assumption of finiteness of the scalar field. 
We see from (\ref{solu}) and (\ref{R}) that for the minimal coupling,
the finiteness of $\bar{\phi}$ is related to the regularity of the
horizon. The scalar field diverges in the surface $r=r_0$ for $\lambda\ne 1$,
in this case the scalar of curvature has a non-removable singularity,
what confirms that such surface is not a regular one, but it
corresponds to a naked singularity. We will see
that this is the case for some non-minimal couplings also. 
To this end, let us consider
the Brans-Dicke theory, for which $f(\phi)=\phi$ and 
$h(\phi)=\frac{\omega}{\phi}$. Using the transformation (\ref{ct}) and
(\ref{int}) we can construct its general asymptotically flat, static, and
spherical symmetric solution starting from the minimally coupled solution
$(\bar{g}_{\mu\nu},\bar{\phi})$,
\begin{eqnarray}
\label{bd}
g_{\mu\nu} &=& \phi^{-1}\bar{g}_{\mu\nu} , \nonumber \\
\bar{\phi} &=& \sqrt{\frac{3}{2}+\omega}\int_a^\phi \frac{d\xi}{|\xi|}.
\end{eqnarray} 
The expression for $\bar{\phi}$ is divergent for $a=0$. Also, if we choose
$a>0$, then $\phi$ must be positive too to avoid the singularity. Let us
take the solution $(\bar{g}_{\mu\nu},-\bar{\phi})$ of (\ref{el2}), and 
consider $\phi \in [a,\infty)$, $a>0$. In this case we have
\begin{equation}
\label{bd1}
\left( \frac{\phi}{a}\right)^{\sqrt{\frac{3}{2}+\omega}} = 
\left( \frac{r+r_0}{r-r_0} \right)^{\sqrt{2(1-\lambda^2)}}.
\end{equation}
The expression (\ref{bd1}) hides a subtleness in the limit of large
$\omega$, maybe the most important limit in Brans-Dicke theory;
recent solar system experiments has been established $\omega >600$\cite{wi}.
In the limit $\omega\rightarrow\infty$, the left-handed side of
(\ref{bd1}) can be $1$ or $\infty$, according to if $\phi=a$ or $\phi>a$.
Due to the fact that the right-handed side is bounded for any
$\lambda$ and for $r>r_0$, the consistency of the equation imply that
$\lambda$ must be $1$ and $\phi=a$ in the limit $\omega\rightarrow\infty$.
This would guarantee that one gets the General Relativity in the limit
$\omega\rightarrow\infty$. Taking this into account 
we have from (\ref{bd})
\begin{eqnarray}
\label{bds}
\phi &=& a {\cal R}^{-k}, \nonumber \\
ads^2 &=& -{\cal R}^{2\lambda + k} dt^2 +
\left( 1 - \frac{r_0^2}{r^2}\right)^2 {\cal R}^{-2\lambda + k}
\left( dr^2 + r^2 d\Omega^2\right),
\end{eqnarray}
where $k=\sqrt{\frac{4(1-\lambda^2)}{3+2\omega}}$. 
The two-parameter $(\lambda, r_0)$ 
family of solutions (\ref{bds}) corresponds to the general 
asymptotically flat, static, and spherically symmetrical solution of
the Brans-Dicke theory.

Our theorem states that the only black-hole solution of (\ref{bds})
with finite $\phi$ 
is the Schwarzschild one, but, at first sight, we can think that
the null-surface $r=r_0$ might be a horizon for some $\lambda$ or
$\omega$. We can check that such surface is not a regular null-surface,
but instead it is a naked singularity for any solution with
non-constant $\phi$. To this end, consider the scalar of curvature
obtained from (\ref{el1})
\begin{equation}
\label{sbd}
R = \frac{\omega}{\phi^2}g^{\mu\nu}\partial_\mu\phi\partial_\nu\phi
=\frac{4r_0^2 r^4}{(r+r_0)^{4+2\lambda -k}} \times
\frac{\omega k^2}{(r-r_0)^{4-2\lambda +k}}.
\end{equation}  
One has that $4-2\lambda +k>0$ for $\lambda \in [0,1]$ and for
$\omega\in [0,\infty)$, and thus (\ref{bds}) has a non-removable 
singularity for any $\lambda\ne 1$ and $\omega\ne 0$. We see
that the only true black-hole solution is that one for which
$\lambda=1$, {\em i.e.} again the Schwarzschild solution
with $\phi =a$, as it was predicted by the theorem. The case
$\omega=0$ can be ruled out by analyzing the singularities of
quadratic invariants, as for example $R_{\mu\nu}R^{\mu\nu}$, that
can by written through (\ref{el1}) by means of $\phi$. We notice that
the first ho-hair theorem for Brans-Dicke theory is due to 
Hawking\cite{H}, and that  
Bekenstein also proved recently the absence of scalar hairs in
Brans-Dicke theory by using his novel no-hair theorem for
minimally coupled scalar fields with non-quadratic Lagrangian\cite{B1}.

We can extend this result for theories such that $\omega(\phi)$ 
is a $C^1$ function and 
$\lim_{\phi\rightarrow\infty}\omega(\phi)=\omega_{\rm c}$. For such a
case, we can evaluate an asymptotic expression for the scalar of
curvature valid for the vicinity of the horizon, and it will lead us
to the conclusion that the only black-hole solution also for this case
is the Schwarzschild one. From (\ref{int}) one can see that for
$\lim_{\phi\rightarrow\infty}\omega(\phi)=\omega_{\rm c}$ and 
$r\rightarrow r_0$ we have
\begin{equation}
\label{bd2}
\phi(r) \approx a 
{\cal R}^{-\sqrt{\frac{4(1-\lambda^2)}{3 - 2\omega_{\rm c}}}}
\end{equation}
From (\ref{bd2}) we have that the expression for $R$ valid for
$r\rightarrow r_0$ is the same one of (\ref{sbd}), from which
we conclude that there is no scalar hair in the model
of Bergman, Wagoner and Nordtved with 
$\lim_{\phi\rightarrow\infty}\omega(\phi)=\omega_{\rm c}$. The result
is valid for any space-time dimension $n>3$.

We can easily apply analogous arguments 
to prove de absence of scalar hair
in dilaton gravity for any space-time dimension $n>3$.

\section{Final remarks}

In spite of the theorem's broad assumptions, there are situations that
it does not cover. In situations where the divergence of the scalar field
is not related to a naked singularity it is possible, in principle,
to exist a scalar hair.
This is the case of the Bekenstein conformal scalar hair\cite{b}, that
obviously escapes from the theorem's assumptions due to the divergence
of the scalar field in the horizon. Such divergence
is not related to any space-time singularity, and for an observer that
does not interact directly with the scalar field the divergence is
physically harmless.

A recent result due to Zannias\cite{Z} also stresses the relevant role
of the divergence of the scalar field in the existence of hairs.
In our approach, the finiteness of $\phi$ guarantees that the only
null-surface of $g_{\mu\nu}$ corresponds to $r=r_0$. If $\phi$ diverges
for some point of the space-time, say $r_1$, the conformal factor 
$\Omega(r_1)$ in (\ref{ct}) vanishes and consequently
$g_{00}(r_1)=0$, what would induce another null-surface for
$r=r_1$. This is precisely what happens with the Bekenstein conformal
hair. However, in principle 
one can find out case by case asymptotic
expressions for the geometrical quantities, as we did in the Sect. \ref{ttt},
and to control the regions very close to the horizons.

We finish noting that two recent works are devoted to problems similar
to the ones discussed here. In \cite{He} Heusler studies with great detail
the case of self-gravitating nonlinear sigma models, for which the action
would be given in our notation by
\begin{equation}
\label{AH}
S[g,\phi^i] = \int d^4x\sqrt{-g} \left\{ 
R - h_{jk}(\phi^i)g^{\mu\nu} \partial_\mu\phi^j \partial_\nu\phi^k + 
W(\phi^i)
\right\},
\end{equation}
where $i\in (1,\dots,N)$. He proved that the only asymptotically flat,
static, and spherically symmetric black-hole solution of (\ref{AH}) is
the Schwarzschild one. Sudarsky\cite{DS} considered the case where 
$h_{jk}(\phi^i)=\delta_{jk}$, getting the same result in a simpler way.
These results are in agreement with our theorem since the case $N=1$
and $W(\phi)=0$ corresponds to our $f=1$ case. However, we believe that
our proof is much more simpler.

This work was supported by CNPq, under process number 201630/93. The
author is grateful to an anonymous referee for pointing out 
references \cite{He} and \cite{DS}.

\end{document}